\documentclass{article}

\usepackage[margin=1in]{geometry}
\usepackage{graphicx}
\usepackage{siunitx}
\usepackage{subcaption}
\usepackage{caption}
\usepackage{longtable}
\usepackage{lineno,hyperref}
\usepackage{enumitem}
\usepackage{authblk}
\usepackage{pdflscape}

\title{Introspecting the Happiness amongst University Students using Machine Learning}
\author[1]{Sakshi Ranjan \thanks{Corresponding Author \\ Accepted at Happiness Meet-2022, IIT Kharagpur }} % \authfn{1}
\author[1]{Pooja Priyadarshini}
\author[2]{Subhankar Mishra}

\affil[1]{Department of Computer Science, Utkal University, Bhubaneswar-751004, India. sakshi.ranjan07@gmail.com}
\affil[2]{School of Computer Sciences, National Institute of Science Education Research, Bhubaneswar-752050, India. \\ Homi Bhabha National Institute, Anushaktinagar, Mumbai - 400094, India. smishra@niser.ac.in}

\begin{document}
  \maketitle

\begin{abstract}
Happiness underlines the intuitive constructs of a specified population based on positive psychological outcomes. It is the cornerstone of the cognitive skills and exploring university students' happiness has been the essence of the researchers lately. In this study, we have analyzed the university students' happiness and its facets using statistical distribution charts; designing research questions. Furthermore, regression analysis, machine learning, and clustering algorithms were applied on the world happiness dataset and university students' dataset for training and testing respectively. Philosophy was the happiest department while Sociology the saddest; average happiness score being 2.8 and 2.44 respectively. Pearson coefficient of correlation was 0.74 for Health. Predicted happiness score was 5.2 and the goodness of model fit was 51\%.
train and test error being 0.52, 0.47 respectively. 
On a Confidence Interval(CI) of 5\% p-value was least for Campus Enviroment(CE) and University Reputation(UR) and maximum for Extra-curricular Activities(ECA) and Work Balance(WB) (i.e. 0.184 and 0.228 respectively). RF+Clustering got the highest accuracy(89\%) and Fscore(0.98) and the least error(17.91\%), hence turned out to be best for our study.

% Please include a maximum of seven keywords
% \keywords{Natural Language Processing, Machine Learning, Deep Learning, Google app reviews, Opinion Mining, Sentiment Analysis.}
\end{abstract}
  
  \section{Introduction}
\textit{"Happiness is not a state to arrive at, but a manner of traveling." - Margaret Lee Runbeck.}
Happiness is a fuzzy or floating or a fluctuant concept that has been explicated by researchers(Economics and Psychology) throughout the ages.  It is the cornerstone of cognitive skills. The present-day psychology - Affirmative Psychology has pinpointed psychological health and well-being since the late 1990s\cite{1}. Ed Diener\cite{2}, one of the pioneers in the area of psychological research has set forth that human well-being is closely associated with happiness in the context of modern psychology. Happiness is tagged with manifold outcomes namely, broadened attention, efficiency, innovations, social factors, and health\cite{3}. %Happiness has flourished extensively in research field of Economics and Psychology.
During the past decades, researchers have surveyed several facets of happiness namely, income, economic growth, escalation, human development index, institutions, expenditure, unemployment, and globalization\cite{4}. Ahn et al.\cite{5} work threw light on the national well-being of people based on GDP, %and highlighted other parameters for attaining happiness specifically, 
physical and mental health, good governance, crime, family and social relationships, economy, and corruption. %Using such factors we can precisely answer a question, "What is the happiness score of a country?" or "Which is the happiest country based on happiness ranks?" or "How is the country performing?" 

The problem statement of the study can be picked out as follows. Firstly, despite the exponential growth in research on happiness and the relationships among the factors, scholars are finding it difficult to define the term happiness exactly\cite{6}. %For instance, one person might be happy with a high income and governance, while another without a high income, but with a good health. 
%In context with expressiveness, appraisals and lifestyles happiness still remains a fuzzy terminology with existing debates amongst researchers. 
The vocabularies namely, happiness, psychological sentiments, well-being, and inter-disciplinary satisfaction are closely interrelated and prejudiced, thereby complicating the results of empirical analysis\cite{7}. Secondly, the ample happiness dataset is available for research, the manual calculation is extremely troublesome. Thirdly, in this fast-paced and technology-driven world, interpersonal relationships have deteriorated drastically\cite{8}. %People are unable to express their emotions openly with peers. 
%People lack time to ask and answer questions like "How was your vacation?" or "How happy are you with the college life?" or "How satisfied are you with your work life?" or "Do you like the current government?"

One of the motivations to conduct this study was to overcome the research gap between the general public and the students\cite{9, ss, ss1}. Students' happiness is directly proportional to university rankings as well. University students' happiness can be enumerated by the criterion namely, age, gender, university reputation, time management, work balance, etc\cite{10}. We get to see the world's happiness rankings every year through the Gallup World polls highlighting the overall world population. However, education linked with happiness has caught the attention of researchers lately.

In our study, we collected  280  university students' reviews via a local survey for conducting empirical analysis, department-wise at Utkal University, India. We incorporate a step-by-step approach to perceive the contribution of our paper :
\begin{itemize}

\item Several Research Questions(RQ) encircling happiness were designed; evaluated on the university students' dataset; answered  through visualization using statistical charts and make an intuitive judgment.

\item In-depth analysis of statistics on our survey data from Utkal University, India, and comparing them with the outcomes of the study of Finland\cite{17} and Australia\cite{18}.  
%to find the mean, Standard Errors(SE), coefficients, Mean Squared Error(MSE) and Root Mean Squared Error(RMSE), Standard Deviation(SD), F-statistics, and p-value of 

\item Linear regression and multiple regression were applied on the 2019 world happiness corpus and used students' happiness dataset for testing. Also, happiness facets namely, Freedom, Social, and Health from the world dataset were used to predict university students' happiness scores. 

\item Machine learning (LR, KNN, Random Forest(RF), SVM) and clustering algorithms were used for empirical analysis and compute performance; comparison graphs were picturized.

\end{itemize}
This paper has been drafted into five sections. Section 2 underlines the background of happiness. Section 3 organizes the techniques utilized in the paper. Section 4 tells about the empirical analysis and results obtained. Finally, Section 5 pinpoints the conclusions of the work and upcoming developments.

\section{Related Works}
 In context with the happiness of university students, Frey and Stuzer\cite{8} argued contribution of age to happiness was negligible. In another instance of age group, Blanchflower and Oswald's\cite{10} research studies revealed that youngsters of the US and Europe were dissatisfied and seemed unjustified due to growth in education.  %Their statistics explained that the European students perusing higher education after 18 years  and below 30 years were rewarded at max instead of happiness.  
% however, Esa\cite{17} found age to be an important explanatory variable.  
The study of Esa\cite{17} showed significant results when happiness results of the countries namely, Finland, Sweden, Norway, Spain, Germany, and Mexico were combined;  but the results were found to be negative. They inferred older Finns are less happy as compared to youngsters, 18 and 28-year-old respondents. Clark and Oswald\cite{19} reached similar inferences. People with higher education were dissatisfied with work. They were to explain whether the university students are happy or not. In the research of Diener et al.and Argyle\cite{d2}, there was a minor association between education and happiness based on the top-level of qualification attained. The pleasure from education ensues from work, earnings, and occupations. Hartog and Oosterbeek\cite{20} revealed schooling period was happy and satisfying as compared to happiness at the university. Diener et al.\cite{2} devised the SWLS to rate critical facets of prosperity and happiness. For instance, reviews like "I like my job culture and I am satisfied with the income as well." are answered on a 7-point Likert scale(1 denotes strongly disagree; 7 denotes strongly agree). 
Analyzing happiness amongst people is mainly a statistical approach with minimal algorithm usage. The exemplary results of Esa \cite{17} and Chan et al\cite{18} was one of the strongest reason to follow them for our study. The idea was to analyze happiness amongst the students of Jyvaskyla University, Finland, and Australia respectively. %The survey data so obtained comprised of components influencing happiness and delight. They inferred student's level of joy was highly influenced by WB, TM, and RF and raised issues on Satisfaction. 
They suggested extending the study to other countries and introspecting whether the factors influencing the happiness of Australian and Finnish students hold good for data crawled from across the globe. So, our work emphasizes the unanswered research questions addressed in Moshe\cite{26}. It highlights the supervised and unsupervised algorithmic approach and their comparison via graphs; aggregating a new dataset to predict happiness parameters specifically for university students that were missing in the literature. Table-1 shows the comparative study of existing literature.%Keeping this school of thought in mind, we constructed our study in India in similar fashion.

%\begin{landscape}
\begin{table}[]
\centering
\caption{Comparative Study of Existing Literature.}
\label{tab:my-table}
\begin{tabular}{|llll|llll|}
\hline
\multicolumn{4}{|l|}{\begin{tabular}[c]{@{}l@{}}a. Descriptive Statistics\\  of Respondents.\end{tabular}} & \multicolumn{4}{l|}{\begin{tabular}[c]{@{}l@{}}b. Students' Satisfaction \\ with University life-Overall I'm \\ happy with my university life.\end{tabular}} \\ \hline
Parameter & Chan & Esa & \begin{tabular}[c]{@{}l@{}}Our \\ Study\end{tabular} & Parameter & Chan & Esa & Our study \\ \hline
Sample Size & 749 & 246 & 280 & \begin{tabular}[c]{@{}l@{}}Strongly agree\\ (or Very happy)\end{tabular} & 13.9 & 22.8 & 47 \\
Female\% total & 50.6 & 53.9 & 63 & Agree (or Happy) & 54.6 & 63.8 & 34.7 \\
\begin{tabular}[c]{@{}l@{}}Directly from\\ high school \% \\ of total\end{tabular} & 74.8 & 27.2 & - & Neutral & 23.1 & 8.5 & 12.2 \\
\begin{tabular}[c]{@{}l@{}}Part time\\ employment \% total\end{tabular} & 48.1 & 35.4 & - & Disagree(or Sad) & 7.2 & 4.9 & 6.1 \\
\begin{tabular}[c]{@{}l@{}}Average \\ working hours\end{tabular} & 11.2 & 10.5 & - & \begin{tabular}[c]{@{}l@{}}Strongly disagree\\ (or Very Sad)\end{tabular} & 1.2 & 0 & - \\ \hline
\end{tabular}
\end{table}
%\end{landscape}

\section{Methodology}
\begin{figure*}
    \centering
    \includegraphics[scale=0.5]{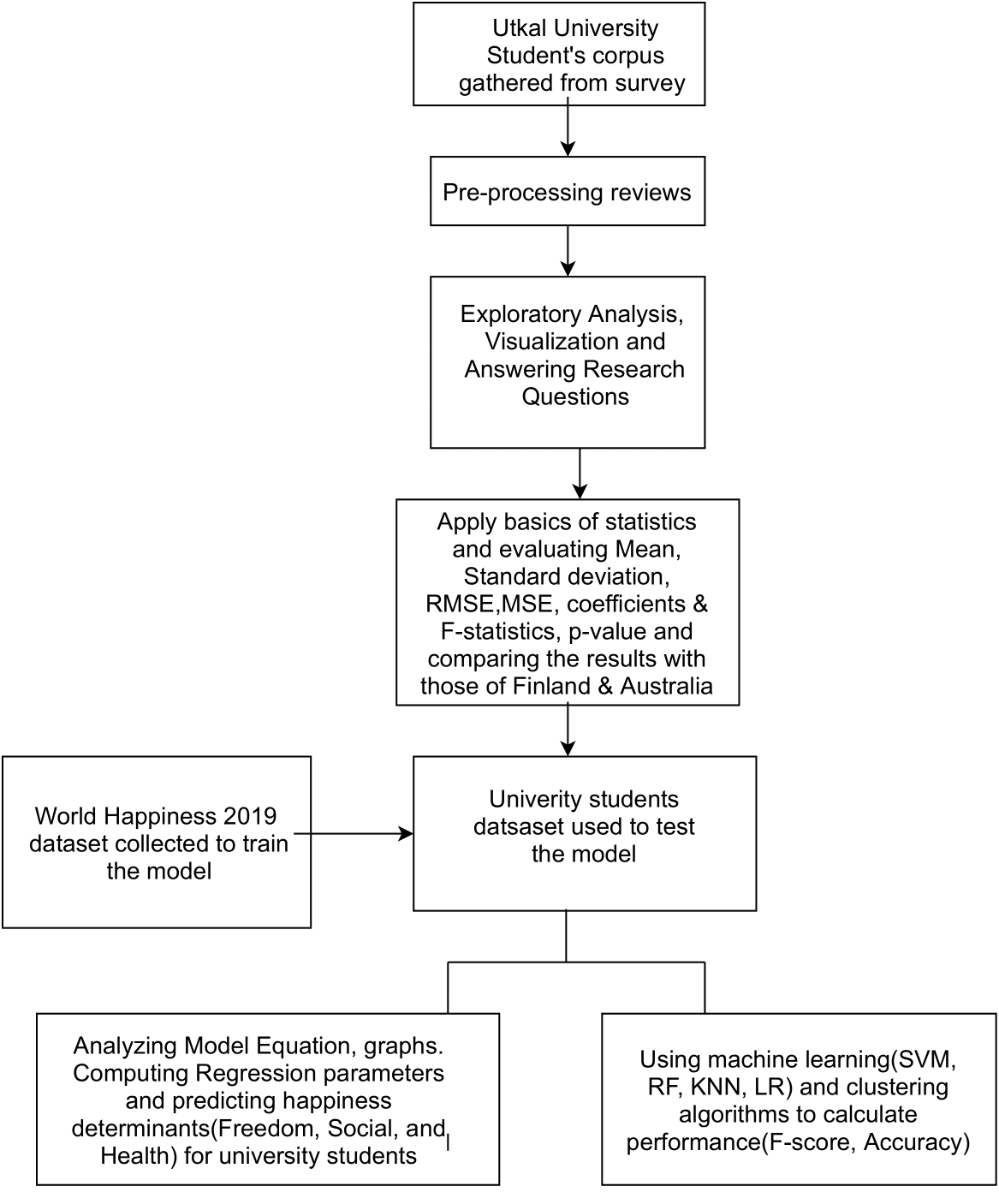}
    \caption{Architectural Design of Student's Happiness.}
    \label{f1}
\end{figure*}

This section briefly describes the techniques used in our study. 
Fig.\ref{f1} explains the architectural design of our study.
\subsection{Data Sources:}

In our study, we used two datasets. Firstly, the corpus, World happiness-2019, was crawled in .csv format.\cite{data}. %It contained 9 columns namely, Overall rank, Country or Region, Score, GDP, Freedom, Corruption, Healthy life, Social Support, and Generosity. 
%Secondly, our line of action was to highlight the happiness of Utkal University(India) students and rank the Department-wise happiness based on the survey data. 
Secondly, we collected a real-time dataset from the university students department-wise(namely, Analytical and Applied Economics, Geography, Odia, Sociology, etc.). The survey was conducted for a month and responses were captured on Google form. It comprised of 10 questionnaires in context with happiness to understand the students' sentiments. They were asked to assign their responses on the designed 4-point Likert scale (1-very happy, 2-Happy, 3-Neutral, 4-Sad). We collected an adequate sample size from among the university students for a comparative research study. The survey conducted was entirely non-compulsory and no baits were offered to gather data. The fields of the survey were Department, University Reputation(UR), Cost of Education(COE), Campus Environment(CE), Good Study Resources(GSR), Relationships Formed(RF), Time Management(TM), Work Balance(WB), Extra Curricular Activities(ECA), Gender, Age Group. Data cleaning and other pre-processing schemes (eliminating lost data, discarding NA values) were initiated on our dataset to enhance the overall efficiency of our model.

\subsection{Quantitative Measures}

\subsubsection{\textbf{Likert Scale}} %It is related to psycho metrics, business and marketing, statistics and used to answer the questionnaires and generate useful insights.
Researchers employ this in a survey to trap the responses of respondents in a close-ended fashion to answer the questionnaires and generate useful insights. Respondents give their level of pleasure or displeasure on a symmetric scale(from 1-5 or 1-7 (1-very happy, 2-happy, 3-neutral, 4-sad, 5-very sad)) of agreement or disagreement. 

\subsubsection{Performance Metrics}
 Precision and recall are combined to give a single criterion termed as F-score. The harmonic mean of precision and recall computes \textbf{F-score}. \textbf{Accuracy} is the ratio of truly predicted observation to the total number of observations.
\begin{equation}
F-score=\frac{2*precision*recall}{precision+recall}
\end{equation}

\begin{equation}
accuracy=\frac{True Positive+True Negative}{True Positive+True Negative +False Positive +False Negative}
\end{equation}

\subsection{Algorithms} - We can term the data as the cornerstone of machine learning algorithms. Such algorithms are trained on the examples by learning from the past experiences and also scrutinizes the former datasets. Training the model on the dataset repeatedly helps to identify the useful patterns from the dataset and generate useful insights. 
It has vast applications in text analysis, image recognition, speech recognition, cognitive science etc. The three types of machine learning algorithms are- Supervised learning, Unsupervised learning, Reinforcement learning.

It is used in the cases where outcome is consecutive and make predictions based on the variables
It helps to design a way to associate the features to the outcomes for prediction. Regression model comprises of unknown terms(beta), error terms(ei), dependent variable(Yi) and independent variable(Xi), the equation is given by(for n data points i=1 to n): %It is used for optimizations, generating insights, supporting decisions, improve miscalculations. Example, predicting value of the stock based on equity and past stocks. The model is trained to find out the estimated price with minimal error. 

\begin{equation}
Y_i=f(X_i,\beta)+e_i    
\end{equation}
  
 \begin{equation}
 f(X_i, \beta)= \beta_0+\beta_i.X_i.  
\end{equation}

\textbf{Linear Regression} A linear relationship is established between the explained and explanatory variables and enables modeling, and is given by: 
It highlights the conditional probability distribution of the dependent variables. One of the shortcomings is over fitting. The equation for Linear Regression is is done to make predictive examination:
\begin{equation}
  Y\hat{a} = bX + A. 
\end{equation}

\textbf{Multiple Regression}- It finds the interrelation between one explained variable and two or more explanatory variable, given by:

\begin{equation}
  Y = \beta_0 + \beta_1.X1 + .......  + \beta_n.X_n. 
\end{equation}

\textbf{Regression Analysis} is the key highlight of our study and highly emphasized.To evaluate the best-fit line for every explanatory variable, Multiple linear regression follows the given steps:1. The regression coefficients for the model are computed by finding out the least model error. 2.The t-statistic for the entire model. 3.The corresponding p-value for the model. 4.The t-statistic and p-value for every regression coefficient are taken into account for the model.Machine learning algorithms are used to make predictions and classifications from the abundant dataset available. For instance: \textbf{SVM} plots the data to an N(no. of features) dimensional space and a hyperplane identify such points during regression and classification. \textbf{RF} utilizes decision trees for training the dataset based on certain attributes. \textbf{KNN} classifies the dataset based on the no. of votes obtained from k nearest neighbors. Aggregation of values from k such neighbors is used as output. \textbf{Clustering}, an unsupervised learning technique, deal with a large, complex, unlabelled datasets. K-means and agglomerative methods were used for our study.

The three-fold experimental demonstrations so performed included - The exploratory analysis of the survey data and answering the RQ's through statistical distribution graphs. The visualization aspect,listed in Table-3, showcased the cognitive skills of students incorporated with fundamentals of statistics. Secondly, a statistical analysis was done on the survey dataset using STATA. The statistical outcomes(RMSE, MSE, Test and Train Error, Mean, SD, Coefficient, F-statistics, p-value etc.) so obtained were compared with those from the study of work of Chan\cite{18} and Esa\cite{17} from university of Australia and Finland. Thirdly, Linear Regression and Multiple Regression was applied on the world happiness-2019 dataset for training the model. For testing we used our survey data of university students. Particularly, we predicted the happiness score for the happiness factors(Freedom, Social, Health) for our test data. Model equations were designed along with computation of regression parameters. Ranking amongst the department was highlighted based on happiness of students. We used scatterplots and modeling plots for making inferences. Then the performance of model was tested using classifiers(LR, KNN, SVM, RF) and clustering algorithms. All graphs were plotted in Datawrapper software available online. A 3D view of the graphs can be derived from the links in reference section of our study.  

%\begin{landscape}
\begin{table}[]
\centering
\caption{Analyzing the Research Questions}
\label{t3}

%\begin{longtable}[c]{p{3em} p{15em} p{5em}  p{20em}}
%\caption{Analyzing the Research Questions}
\begin{tabular}{p{3em} p{10em} p{3em} p{19em}}
Serial & Research Question & Figure & Answer  \\
\hline

RQ1 & What is the distribution of happiness score made department-wise? & 2a &The happiness range plot showed an eminent distinction in Law, Odia and Physics department. On the contrary, Zoology and Chemistry had negligible scores. Average score turned out to be 2.51. \\

%RQ2 &  How is the distribution of happiness factors made based on the 4 happiness classes? & 2b &The plot illustrates students were pleased with CE(averagely 25\%) in all 4 classes; WB and COE showed a remarkable variation in 4 classes(11\%).\\

%RQ3 & What is the overall participation percentage of students based on happiness factors? & 2c &The pie chart outlines the exceptional happiness score for CE based on the overall participation of females(age 25-30) and negligible male participation. Overall percentage resulting to 17\% for RF and ECA. \\

RQ2 &  What is the percentage distribution of happiness classes based on happiness factors? & 2b &The pie chart portrays irrespective of the department, considering all the happiness factors, averagely 46\% of students were Very Happy and Sad while Happy and Neutral resulting in 44\% and 45\%. Least percent(13\%) for Happy and Sad classes.  \\

RQ3 &  What is the departmental statistics based on all happiness factors?& 2c & The plot depicts that average happiness score of 3.3 was seen in Chemistry department for TM, GSR for English, COE for Zoology and WB for Personal Management.\\

%RQ6 &  What is the distribution of happiness factors based on their SD department-wise? & 3b & The plot highlights the maximum standard deviation of Happiness scores(3 and 2.6) in the Botany and Philosophy department in ECA. WB in Biotechnology and Philosophy department is also catching up(2.7 and 2.3).\\

RQ4 & How are the 'Very Happy' students grouped together based on Happiness parameters department-wise? & 2d & The stacked column chart depicts Commerce students were Very Happy with TM; Averagely students of Women Studies, History and Archaeology, Chemistry, Sociology, and Public administration were Very Happy with UR, COE, WB and, ECA respectively.\\

RQ5 & How are the 'Happy' students grouped together based on Happiness parameters department-wise?  & 2e &The stacked column chart depicts English and Commerce students were Happy with WB and COE. Averagely, students from Computer Science and Applications, Political Science, Geography, and Sociology were Happy with WB, CE, GSR respectively.\\

RQ6 & How are the 'Neutral' students grouped based on Happiness parameters department-wise? & 2f &The stacked column chart depicts Pharmacy and Philosophy students were Neutral with RF and COE.\\

RQ7 & How are the 'Sad' students grouped based on Happiness parameters department-wise?  & 2g & The stacked column chart depicts Library and Information Science students were Sad about RF. Averagely, students from Chemistry and Zoology were Sad with TM and COE respectively.\\

RQ8 &  What is the male participation statistics based on university students survey? & 2h &The plot illustrates average happiness score(2.82) for CE and SD(1.41) for ages 30-35. Average happiness score(2.64) for RF and SD(2.12) for ages 25-30.\\

RQ9 & What is the female participation statistics based on university students survey?& 2i & The plot illustrates average happiness score(2.82) for CE and SD(1.41) for ages 30-35. Average happiness score(2.64) for RF and SD(1.41) for ages 25-30.\\
\hline
\end{tabular}
\end{table}
%\end{landscape}

\begin{figure}[!tbp]
  \centering
       \begin{minipage}[b]{0.3\textwidth}
    \includegraphics[width=\textwidth]{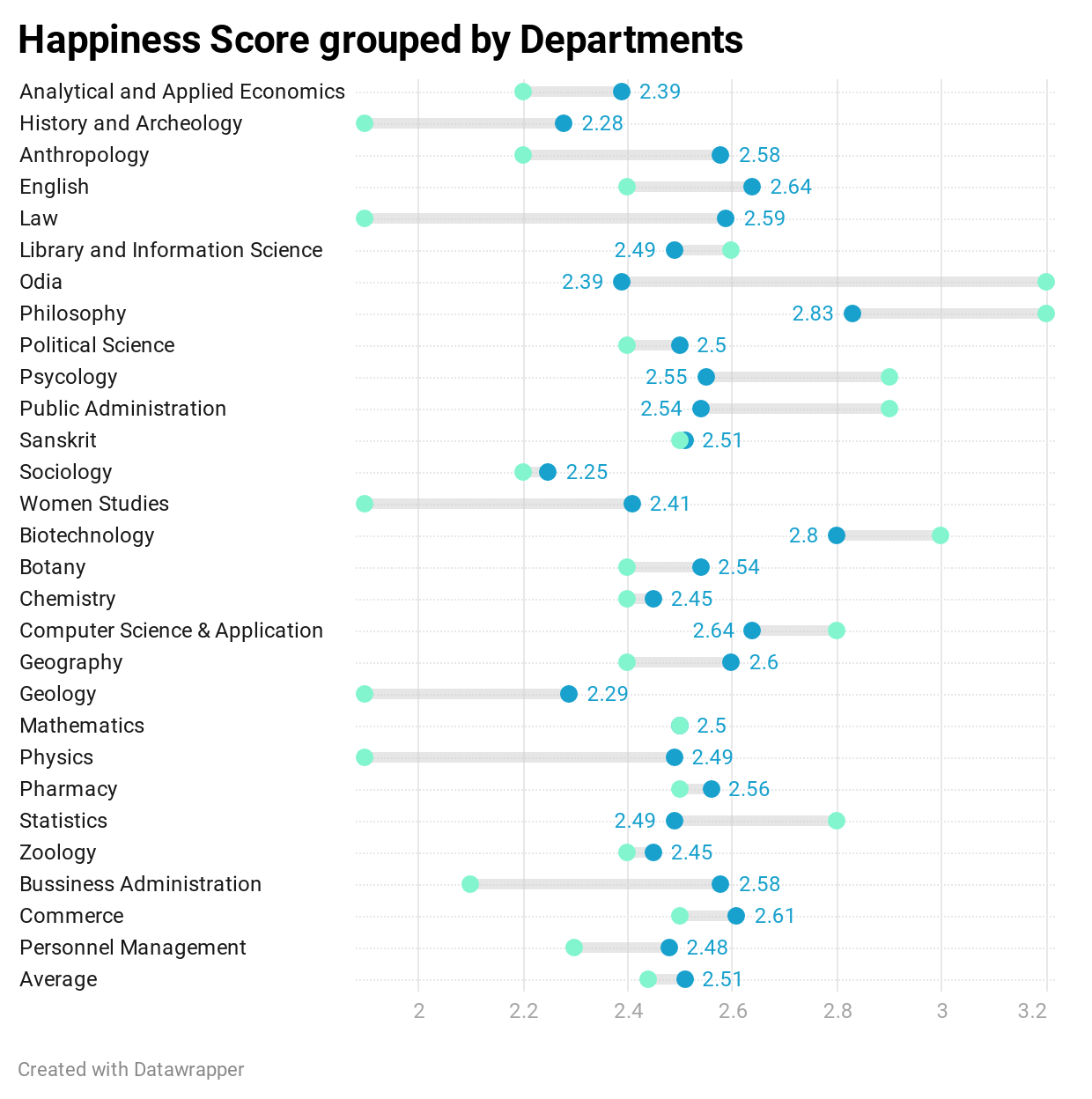}
    \subcaption{}
    \label{rq1}
\end{minipage}
 %\hfill
  %     \begin{minipage}[b]{0.3\textwidth}
   % \includegraphics[width=\textwidth]{RQ11.png}
    %\subcaption{}
    %\label{rq11}
%\end{minipage}
 %\hfill
  %  \begin{minipage}[b]{0.3\textwidth}
   % \includegraphics[width=\textwidth]{RQ13.png}
    %  \subcaption{}
     %    \label{rq13}
 % \end{minipage}
  \hfill
  \begin{minipage}[b]{0.3\textwidth}
    \includegraphics[width=\textwidth]{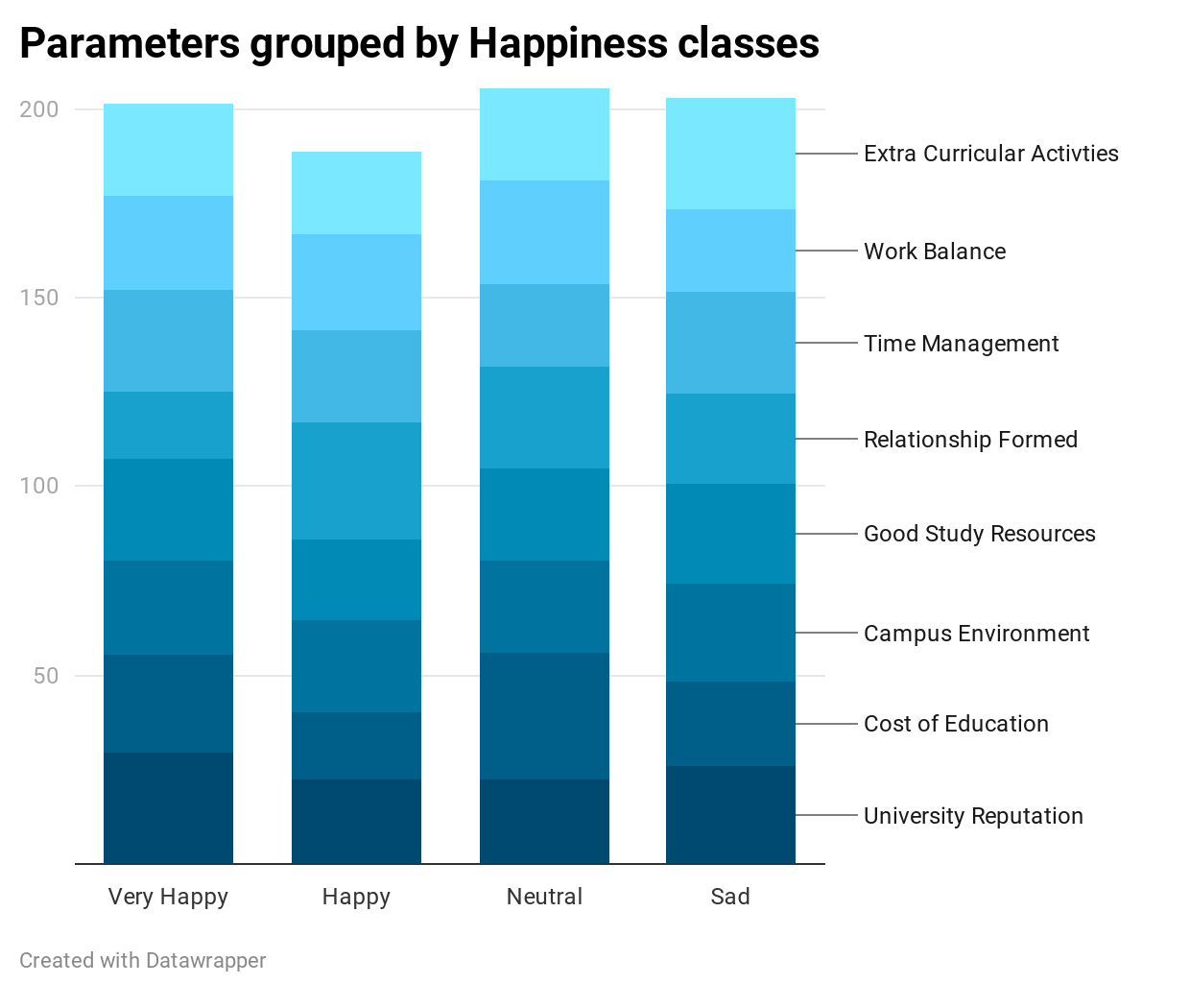}
     \subcaption{}
         \label{rq10}
  \end{minipage}
  \hfill
  \begin{minipage}[b]{0.3\textwidth}
    \includegraphics[width=\textwidth]{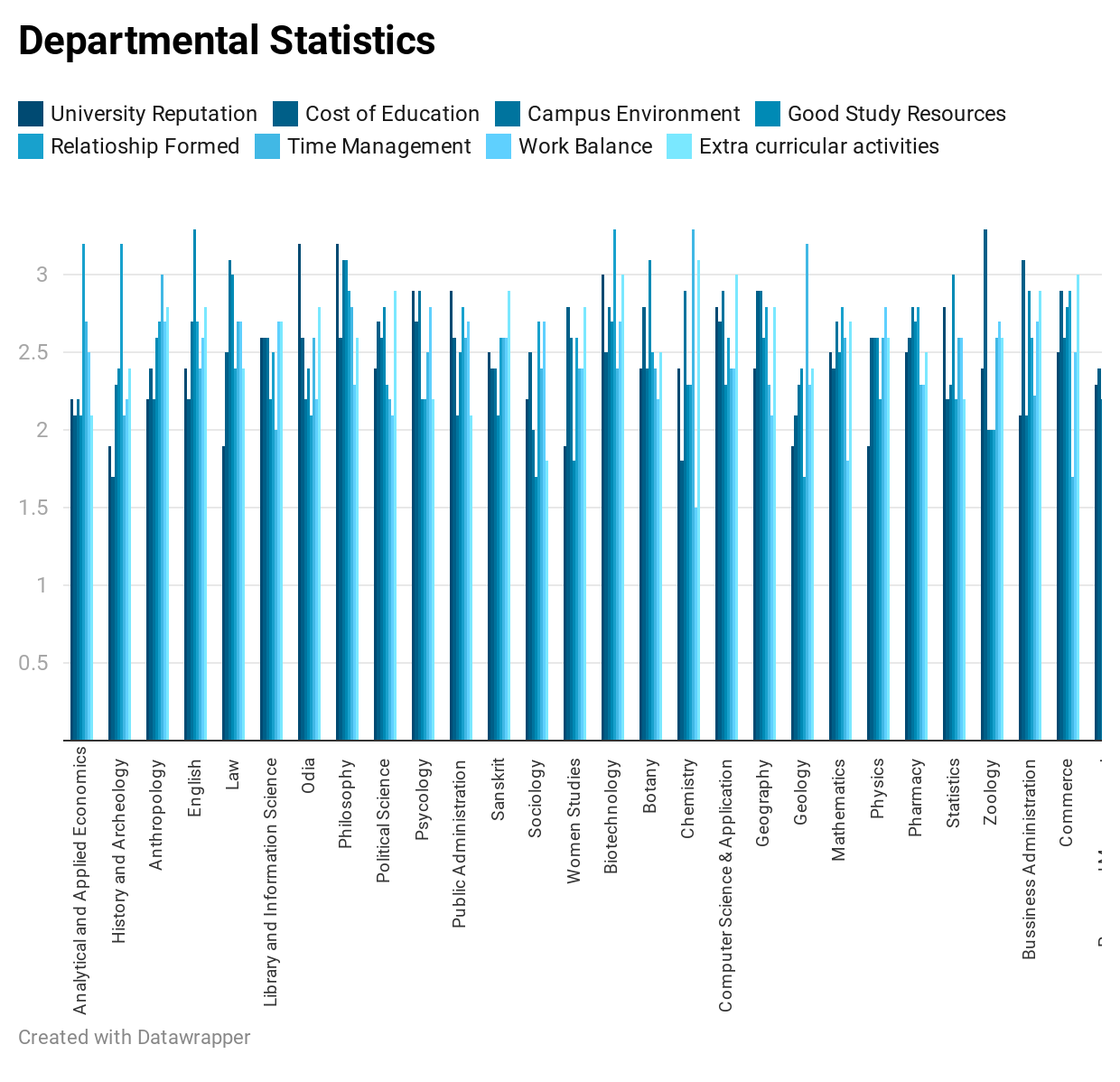}
      \subcaption{}
         \label{rq6}
  \end{minipage}
 \hfill
 % \begin{minipage}[b]{0.3\textwidth}
  %  \includegraphics[width=\textwidth]{RQ12.png}
    %  \subcaption{}
     %    \label{rq12}
 % \end{minipage}
 
  \begin{minipage}[b]{0.3\textwidth}
    \includegraphics[width=\textwidth]{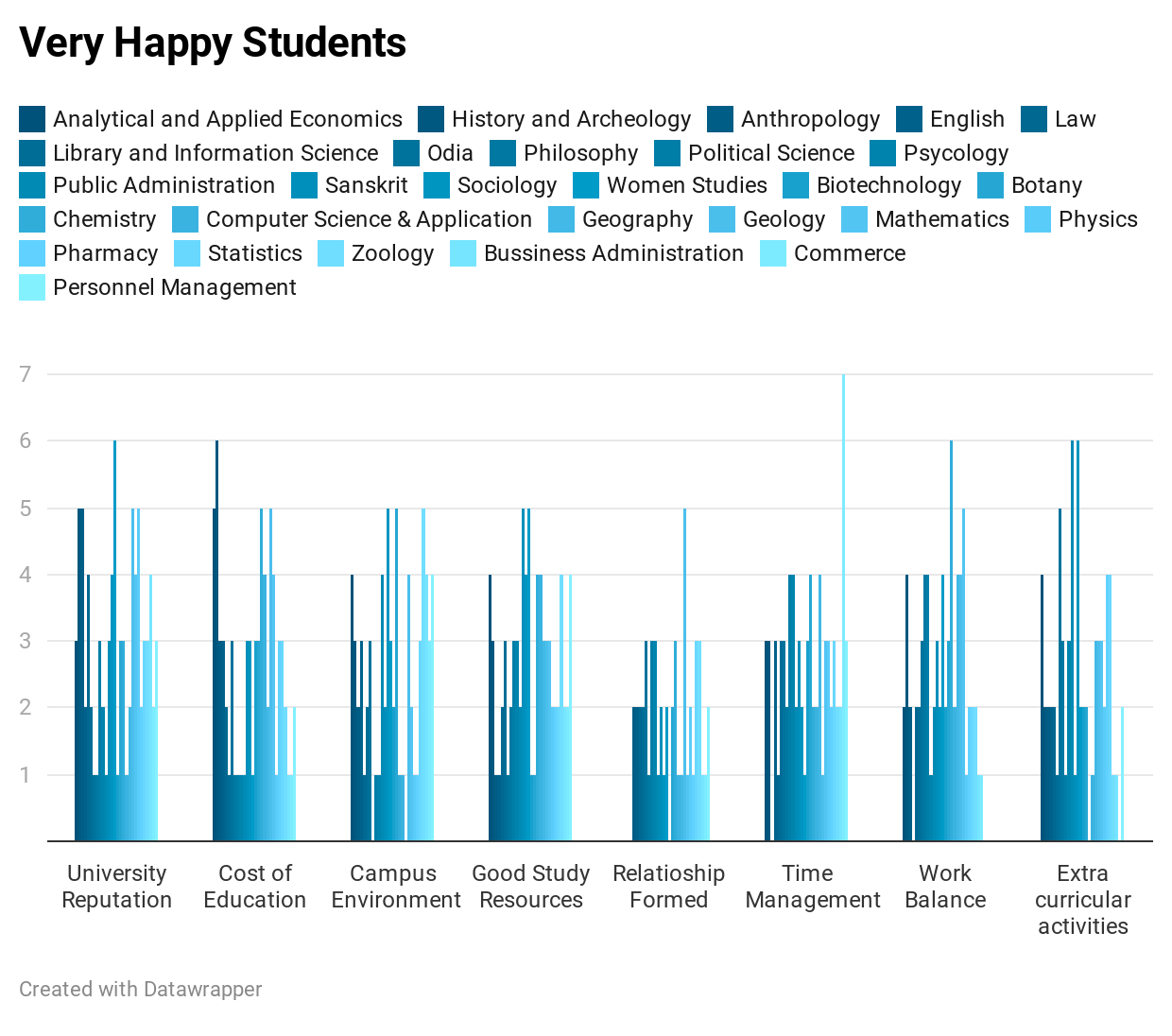}
    \subcaption{}
    \label{rq2}
\end{minipage}
 \hfill
       \begin{minipage}[b]{0.3\textwidth}
    \includegraphics[width=\textwidth]{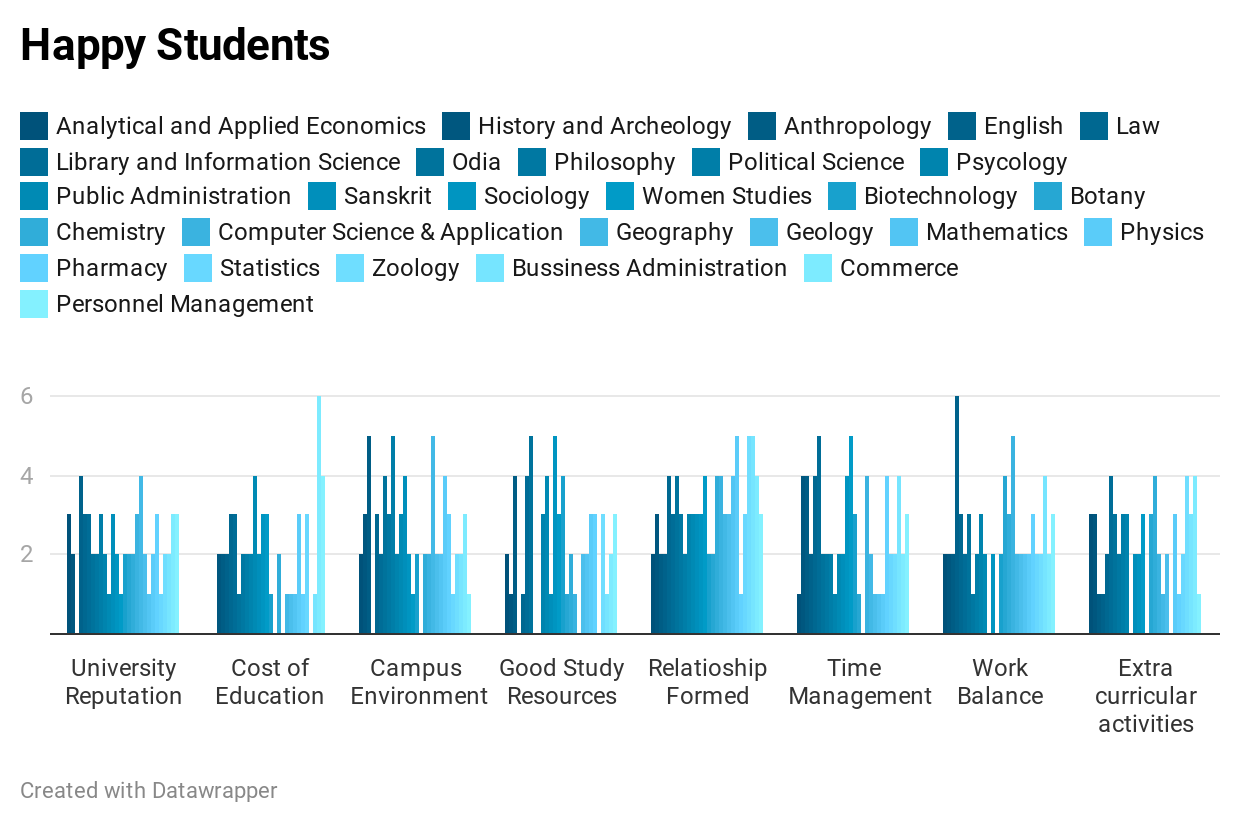}
    \subcaption{}
    \label{rq3}
\end{minipage}
 \hfill
    \begin{minipage}[b]{0.3\textwidth}
    \includegraphics[width=\textwidth]{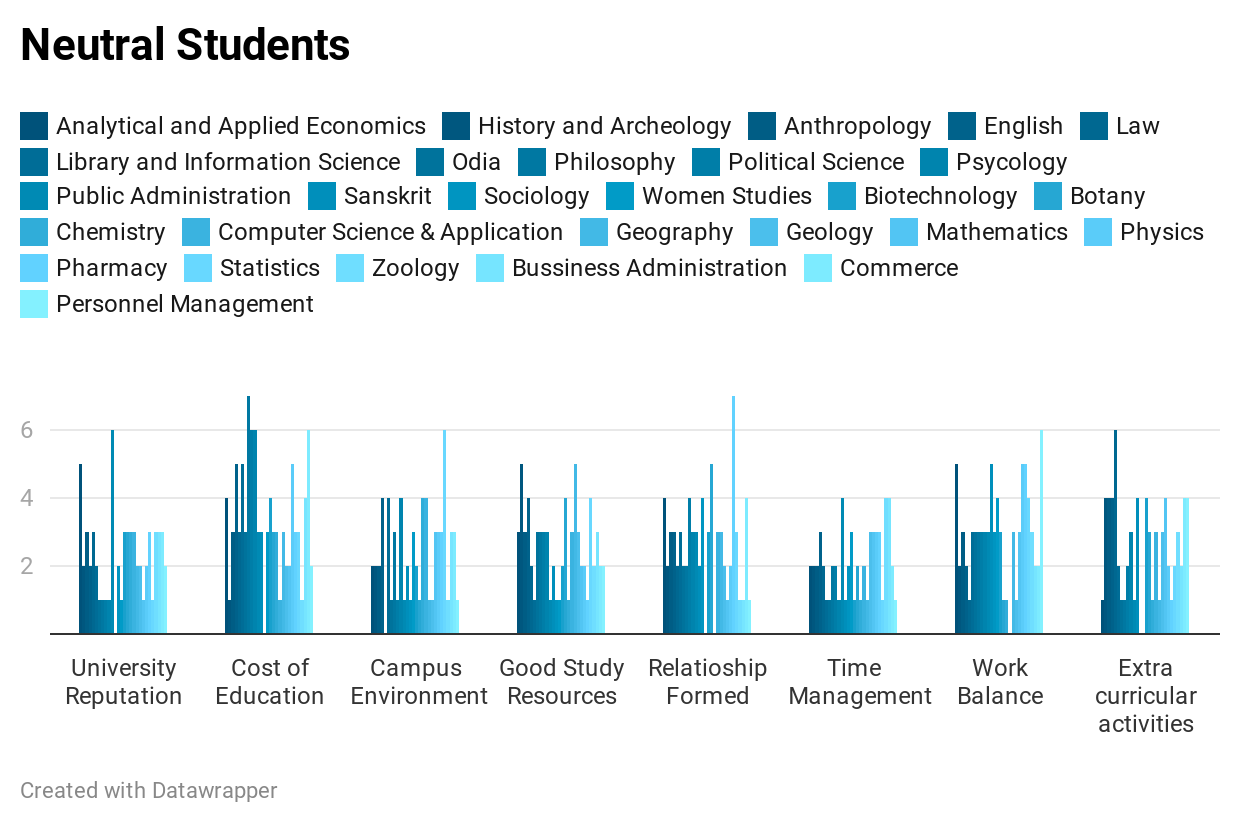}
      \subcaption{}
         \label{rq4}
  \end{minipage}
  \hfill
  \begin{minipage}[b]{0.3\textwidth}
    \includegraphics[width=\textwidth]{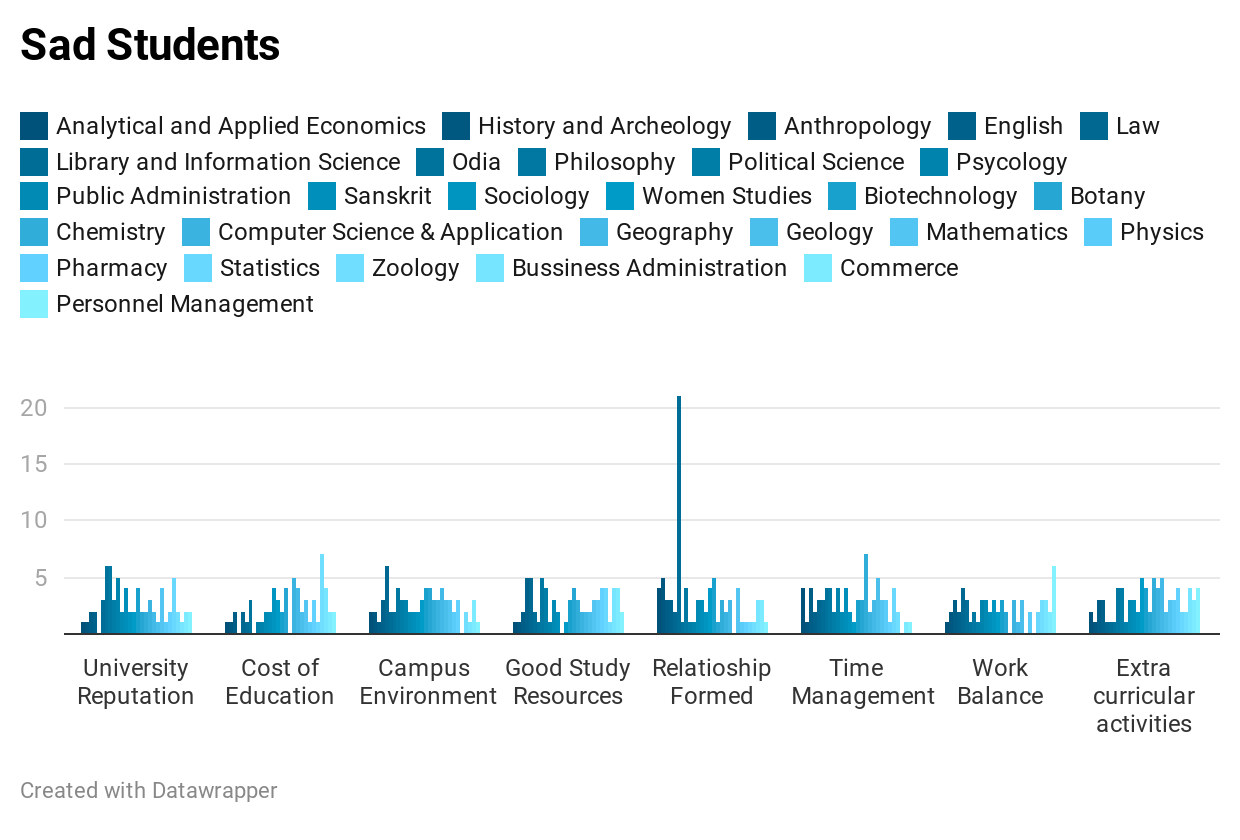}
     \subcaption{}
         \label{rq5}
  \end{minipage}
  \hfill
   \begin{minipage}[b]{0.3\textwidth}
    \includegraphics[width=\textwidth]{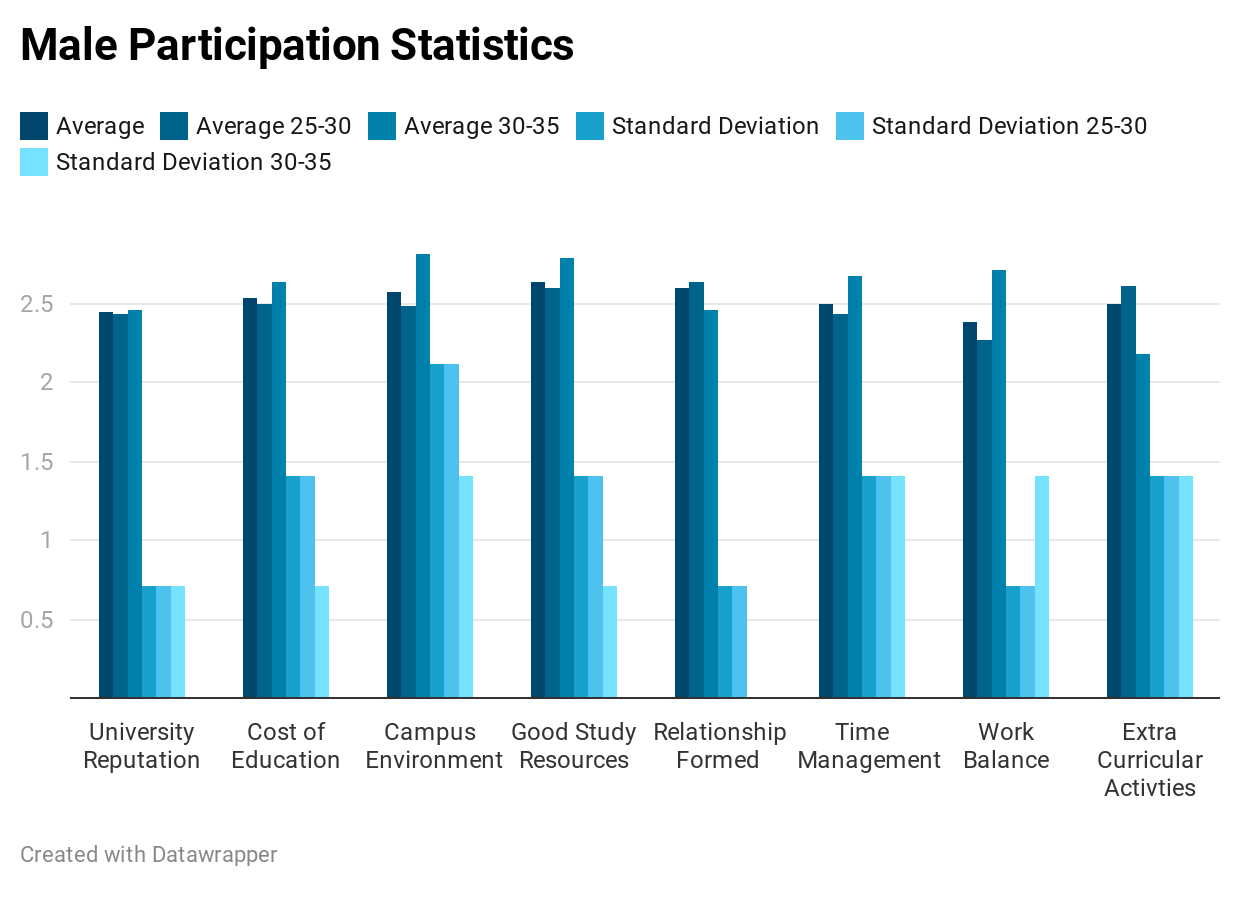}
      \subcaption{}
         \label{rq6}
  \end{minipage}
  \hfill
  \begin{minipage}[b]{0.3\textwidth}
    \includegraphics[width=\textwidth]{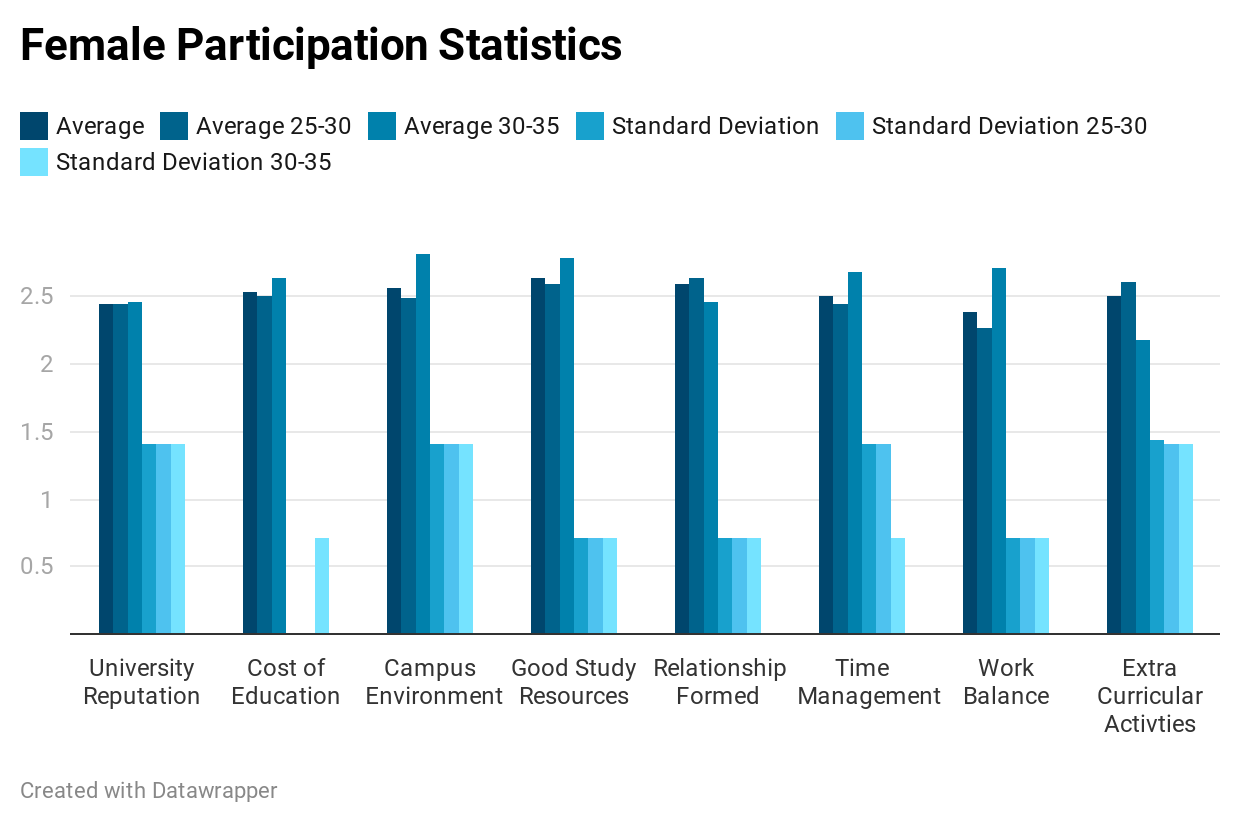}
      \subcaption{}
         \label{rq12}
  \end{minipage}
  \caption{Visualization of Happiness Distribution Plots}
    \end{figure}

%\begin{landscape}
\begin{table}[]
\centering
\caption{Evaluating Regression Analysis Parameters}
\label{t5}
\begin{tabular}{l|l|l|l}
Dep Variable: & score & R-Squared(uncentered): & 0.986 \\
Model: & OLS & Adjusted R-Square(uncentered): & 0.986 \\
Method: & Least Squares & F-Statistic: & 362.2 \\
Date: & Fri, 20 Nov 2020 & Prob (F-statistic): & 7.57e-281 \\
Time: & 20:03:40 & Log-Likelihood: & -307.69 \\
No. of Observations: & 312 & AIC: & 627.4 \\
Df Residual: & 306 & BIC: & 649.8 \\
Df Model: & 6 & Kurtosis: & 3.612 \\
Covariance Type: & nonrobust & Skew: & 0.212
\end{tabular}

\end{table}
%\end{landscape}

%\begin{landscape}
\begin{table}[]
\centering
\caption{Descriptive Statistics of Explanatory Variables}
\label{t5}
\begin{tabular}{p{19em} p{2em} p{13em}}
Variables & & Values \\
\hline
\textbf{Age}: Age of Student (years).& &Mean=20(25-30years); Mean=22.97(30-35years); SD=2.82 \\
\textbf{Gender:} Contrary variable, Male or Female & & Males=37\%; females=63\% \\
\textbf{UR}: estimation of student's satisfaction with the university in which they are studying based on its reputation and overall rankings & & Very Happy=29.64\%, Happy=22.50\%, Neutral=22.14\%, Sad=25.71\% \\
\textbf{COE}: cost incurred by the students for getting higher education from educational institutions. & & Very Happy=26.07\%, Happy=17.86\%, Neutral=33.57\%, Sad=22.14\% \\
\textbf{CE}: an environment that provides structured and regular learning opportunities, good hostel; sports facilities, conditions of security and safety for the students encircling their comfort at college. & & Very Happy=25\%, Happy=25.29, Neutral=24.64, Sad=25.71 \\
\textbf{GSR}: resources like lab equipment, and library books which help the students to achieve their academic and personal goals. & & Very Happy=27.14\%, Happy=21.43\%, Neutral=24.64\%, Sad=26.43\% \\
\textbf{RF}: estimates of the bond formed between students and their faculty. & & Very Happy=17.86\%, Happy=31.07\%, Neutral=27.14\%, Sad=23.93\% \\
\textbf{TM}: estimation of time concerning work balance and university chores, meeting academic deadlines and goals, plentiful recreational time along with academics. & & Very Happy=26.79\%, Happy=24.29\%, Neutral=21.79\%, Sad=27.14\% \\
\textbf{WB}: estimation of how well a student manages his semester works regularly and remains consistent throughout. & &  Very Happy=25\%, Happy=25.36\%, Neutral=27.50\%, Sad=21.79\% \\
\textbf{ECA}: estimation of activities performed by students which fall outside their educational course curriculum. & & Very Happy=24.29\%, Happy=21.79\%, Neutral=24.64\%, Sad=29.29\% \\
\textbf{Freedom: predicted variable}: estimates the privilege of students to openly think, act, and interact with seniors, juniors, and peers. & & Very Happy=1.97\%, Happy=6.47\%, Neutral=4.95\%, Sad=4.23\% \\
\textbf{Health: predicted variable}: estimates the medical facility given to students based on health conditions. & & Very Happy=8.71\%, Happy=9.1\%, Neutral=9.09\%, Sad=9.09\% \\
\textbf{Social: predicted variable}: estimates the social activities like tree plantation, promoting underprivileged students through campaigns and initiatives conducted in university. & & Very Happy=13.6\%, Happy=15.73\%, Neutral=12.11\%, Sad=11.91\% \\
\end{tabular}

\end{table}

%\end{landscape}
\begin{table}[]
\centering
\caption{Descriptive Statistics of Predicted variables}
\label{tab:my-table}
\begin{tabular}{|l|l|l|l|l|}
\hline
\textbf{Parameters}       & \textbf{Social} & \textbf{Health} & \textbf{Freedom} & \textbf{Score} \\ \hline
\textbf{coeff}            & 2.26            & 1.25            & 1.86             & -              \\
\textbf{std error}        & 0.15            & 0.26            & 0.28             & -              \\
\textbf{t}                & 14.78           & 4.8             & 6.44             & -              \\
\textbf{P\textgreater{}t} & 0               & 0               & 0                & -              \\
\textbf{0.025}            & 1.96            & 0.73            & 1.29             & -              \\
\textbf{0.975}            & 2.56            & 1.76            & 2.43             & -              \\
\textbf{count}            & 312             & 312             & 312              & 312            \\
\textbf{mean}             & 1.21            & 0.66            & 0.42             & 5.39           \\
\textbf{SD}               & 0.30            & 0.25            & 0.15             & 1.11           \\
\textbf{min}              & 0               & 0               & 0                & 2.85           \\
\textbf{25\%}             & 1.05            & 0.48            & 0.32             & 4.51           \\
\textbf{50\%}                      & 1.26            & 0.69            & 0.44             & 5.37           \\
\textbf{75\%}                      & 1.45            & 0.85            & 0.54             & 6.17           \\
\textbf{max}                       & 1.64            & 1.41            & 0.72             & 0.76           \\
\textbf{Pearson}                   & 0.65            & 0.74            & 0.55             & -              \\ \hline
\end{tabular}
\end{table}

% %\begin{landscape}
% \begin{table}[]
% \centering
% \caption{Descriptive Statistics of Predicted variables}
% \label{tab:my-table}
% \begin{tabular}{|llllllllllllllll|}
% \hline
% \begin{tabular}[c]{@{}l@{}}Param-\\ eters\end{tabular} & coeff & \begin{tabular}[c]{@{}l@{}}std\\ err\end{tabular} & t & P\textgreater{}t & 0.025 & 0.975 & count & mean & SD & min & 25\% & 50\% & 75\% & max & \begin{tabular}[c]{@{}l@{}}Pear-\\ son\end{tabular} \\ \hline
% Social & 2.26 & 0.15 & 14.78 & 0 & 1.96 & 2.56 & 312 & 1.21 & 0.3 & 0 & 1.05 & 1.26 & 1.45 & 1.64 & 0.65 \\
% Health & 1.25 & 0.26 & 4.8 & 0 & 0.73 & 1.76 & 312 & 0.66 & 0.25 & 0 & 0.48 & 0.699 & 0.85 & 1.41 & 0.74 \\
% \begin{tabular}[c]{@{}l@{}}Freed-\\ om\end{tabular} & 1.86 & 0.28 & 6.44 & 0 & 1.29 & 2.43 & 312 & 0.42 & 0.15 & 0 & 0.32 & 0.44 & 0.54 & 0.72 & 0.55 \\
% Score & - & - & - & - & - & - & 312 & 5.39 & 1.11 & 2.85 & 4.51 & 5.37 & 6.17 & 0.76 & - \\ \hline
% \end{tabular}
% \end{table}
% %\end{landscape}

%\begin{landscape}
\begin{table}[]
\centering
\caption{Estimates of Model of satisfaction with university life.}
\label{tab:my-table}
\begin{tabular}{|llll|ll|ll|}
\hline
Variable & \begin{tabular}[c]{@{}l@{}}Coefficient\\ (Std error)\end{tabular} & y-predict & y\_test & \begin{tabular}[c]{@{}l@{}}Predicted\\  variable\end{tabular} & \begin{tabular}[c]{@{}l@{}}Coefficient\\ (Std. error)\end{tabular} & Parameters & Results \\\hline

Age & 0.929(0.551) & 5.459 & 5.124 & Freedom & 0.99(0.525) & Pseudo R-Square & 0.767 \\
Gender & 1.065(3.842) & 4.592 & 4.297 & Health & 0.99(0.696) & Model Fitting & 51.022 \\
UR & 0.881(0.569) & 6.214 & 6.455 & Social & 1(0.425) & MSE & 0.26 \\
COE & 0.971(0.663) & 6.257 & 6.786 &  &  & RMSE & 0.51 \\
CE & 0.520(1.369) & 6.163 & 6.298 &  &  & Intercept & 3.36 \\
GSR & 0.943(0.978) & 4.852 & 3.819 &  &  & Coefficient & 2.25 \\
RF & 1.02(1.22) & 4.391 & 4.633 &  &  & Constant & 1.85 \\
TM & 0.936(0.475) & 3.942 & 4.971 &  &  & rscore & 0.63 \\
WB & 1.404(0.878) & 5.907 & 5.754 &  &  & Train error & 0.528 \\
ECA & 1.447(0.955) & 5.310 & 4.857 &  &  & Test error & 0.74 \\ \hline
\end{tabular}
\end{table}
%\end{landscape}

\begin{figure}[!tbp]
  \centering
       \begin{minipage}[b]{0.3\textwidth}
    \includegraphics[width=\textwidth]{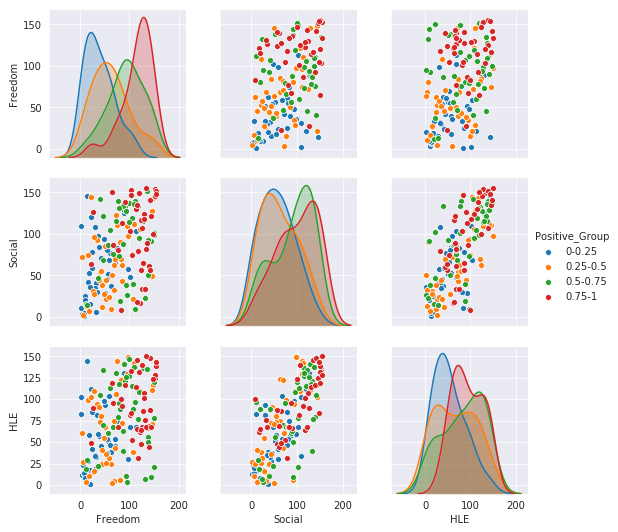}
    \subcaption{Pairwise Modelling plot for predicted variables}
    \label{lr1}
\end{minipage}
 \hfill
       \begin{minipage}[b]{0.3\textwidth}
    \includegraphics[width=\textwidth]{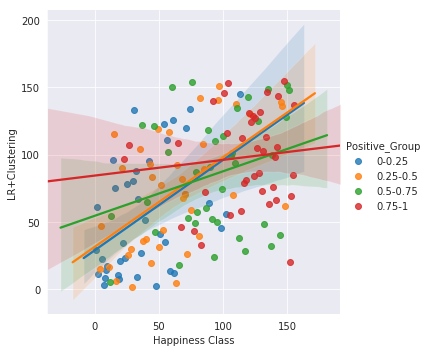}
    \subcaption{LR+Clustering Modelling plot}
    \label{lr2}
\end{minipage}
 \hfill
    \begin{minipage}[b]{0.3\textwidth}
    \includegraphics[width=\textwidth]{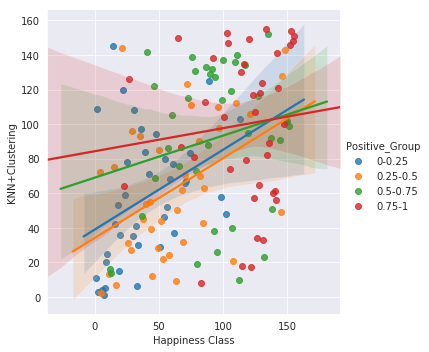}
      \subcaption{KNN+Clustering Modelling plot}
         \label{lr3}
  \end{minipage}
  \hfill
  \begin{minipage}[b]{0.3\textwidth}
    \includegraphics[width=\textwidth]{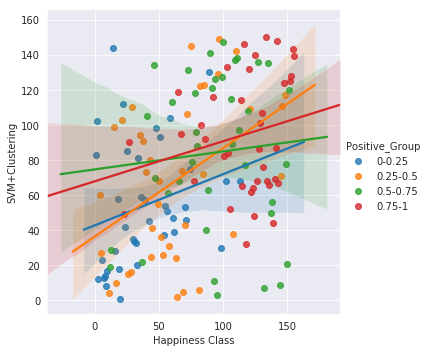}
     \subcaption{SVM+Clustering Modelling plot}
         \label{lr4}
  \end{minipage}
  \hfill
   \begin{minipage}[b]{0.3\textwidth}
    \includegraphics[width=\textwidth]{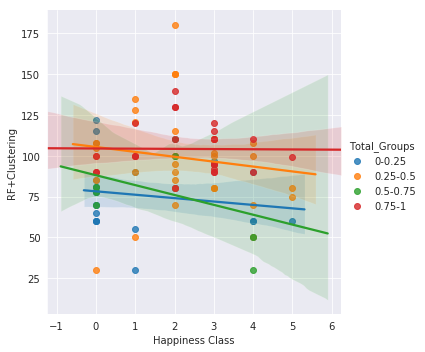}
    \subcaption{RF+Clustering Modelling plot}
    \label{lr2}
\end{minipage}
 \hfill
  \begin{minipage}[b]{0.3\textwidth}
    \includegraphics[width=\textwidth]{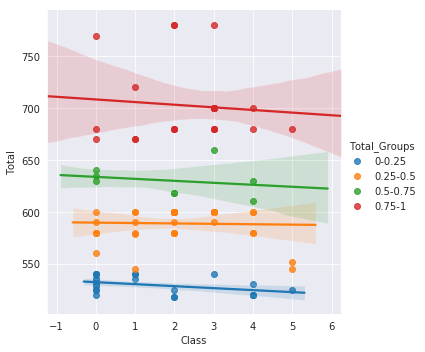}
    \subcaption{Overall Clustering plot}
    \label{lr2}
\end{minipage}
 \hfill
 \caption{Visualization of Modelling based on Happiness Classes }
    \end{figure}

\begin{figure}[!tbp]
  \centering
       \begin{minipage}[b]{0.3\textwidth}
    \includegraphics[width=\textwidth]{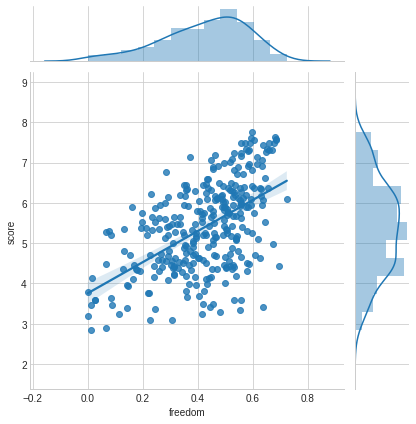}
    \subcaption{Scatter-plot for Happiness Score and Freedom}
    \label{m1}
\end{minipage}
 \hfill
       \begin{minipage}[b]{0.3\textwidth}
    \includegraphics[width=\textwidth]{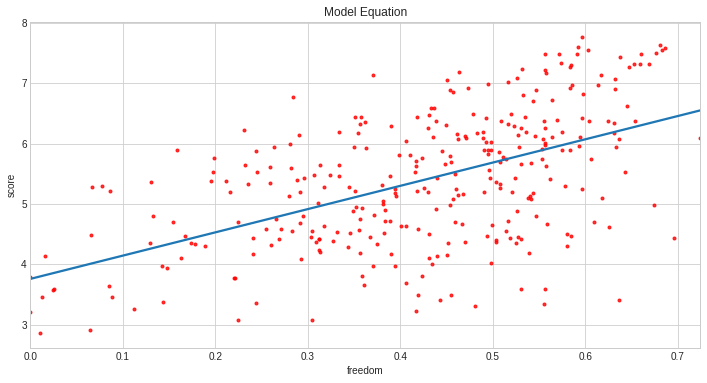}
    \subcaption{Modeling for Happiness score vs Freedom}
    \label{m2}
\end{minipage}
 \hfill
    \begin{minipage}[b]{0.3\textwidth}
    \includegraphics[width=\textwidth]{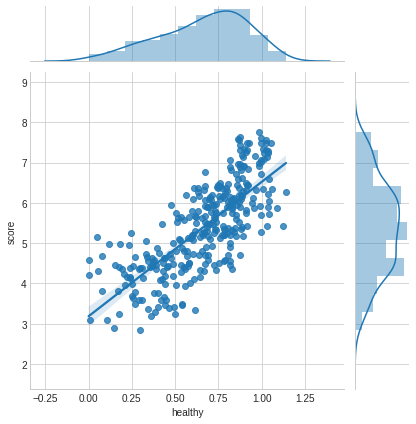}
      \subcaption{Scatter-plot for Happiness Score and Health}
         \label{m3}
  \end{minipage}
  \hfill
  \begin{minipage}[b]{0.3\textwidth}
    \includegraphics[width=\textwidth]{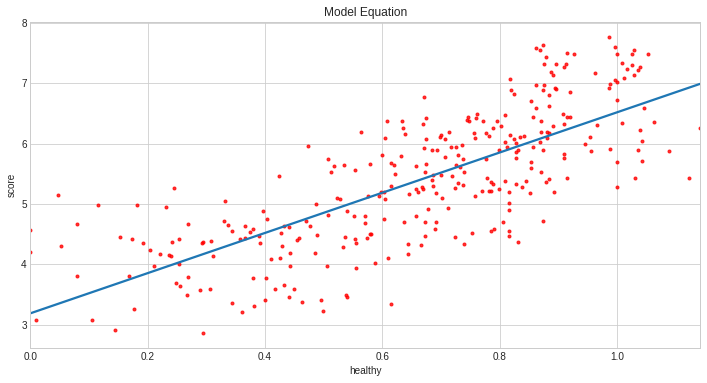}
     \subcaption{Modeling for Happiness score vs Health}
         \label{m4}
  \end{minipage}
  \hfill
    \begin{minipage}[b]{0.3\textwidth}
    \includegraphics[width=\textwidth]{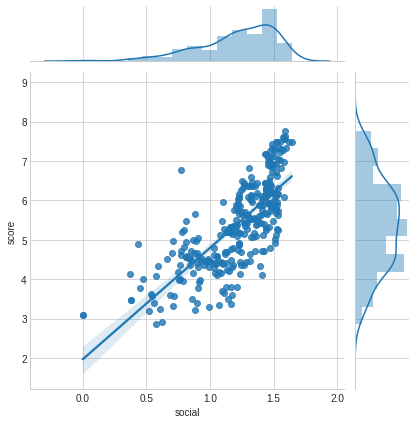}
      \subcaption{Scatter-plot for Happiness Score and Social}
         \label{m5}
  \end{minipage}
  \hfill
  \begin{minipage}[b]{0.3\textwidth}
    \includegraphics[width=\textwidth]{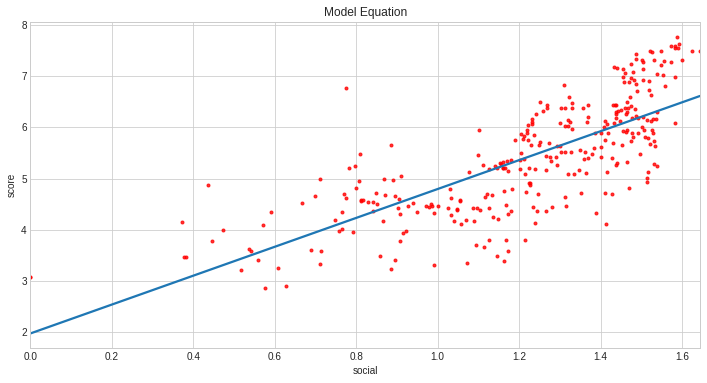}
     \subcaption{Modeling for Happiness score vs Social}
         \label{m6}
  \end{minipage}
  \hfill
  \caption{Visualization of Predicted variables using Regression Analysis for Student’s Happiness}
\end{figure}

    \begin{figure}[!tbp]
       \centering
     \begin{minipage}[b]{0.3\textwidth}
         \centering
         \includegraphics[width=\textwidth]{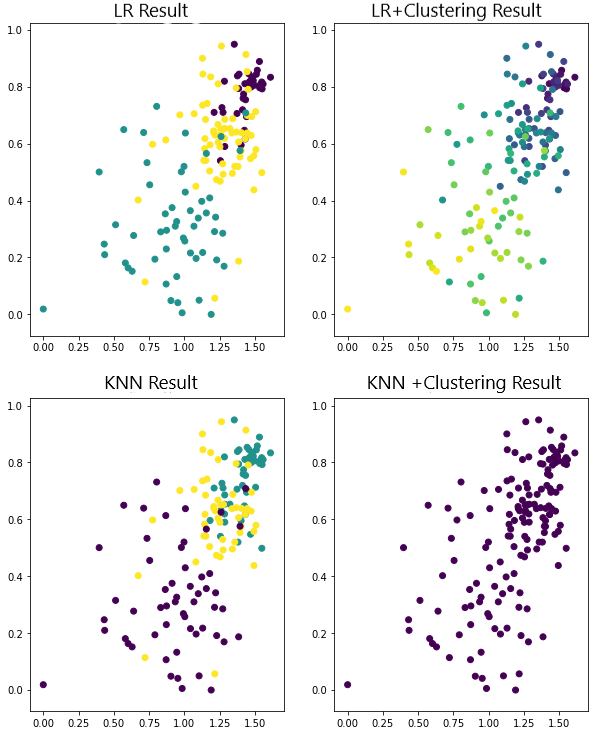}
         %\subcaption{Comparative Plot for Regression parameters}
         \label{rq7}
     \end{minipage}
     \hfill
     \begin{minipage}[b]{0.3\textwidth}
         \centering
         \includegraphics[width=\textwidth]{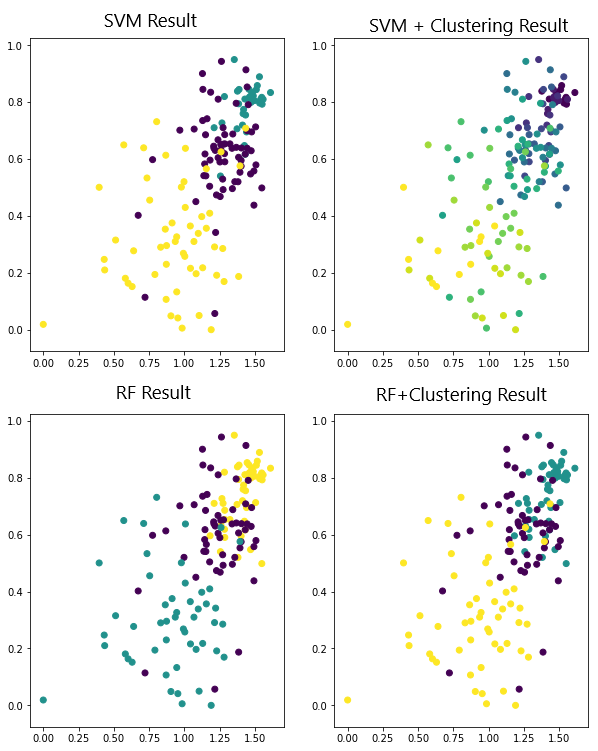}
         %\subcaption{Distribution of Happiness Score after prediction.}
         \label{rq8}
     \end{minipage}
     \hfill
       \begin{minipage}[b]{0.3\textwidth}
         \centering
         \includegraphics[width=\textwidth]{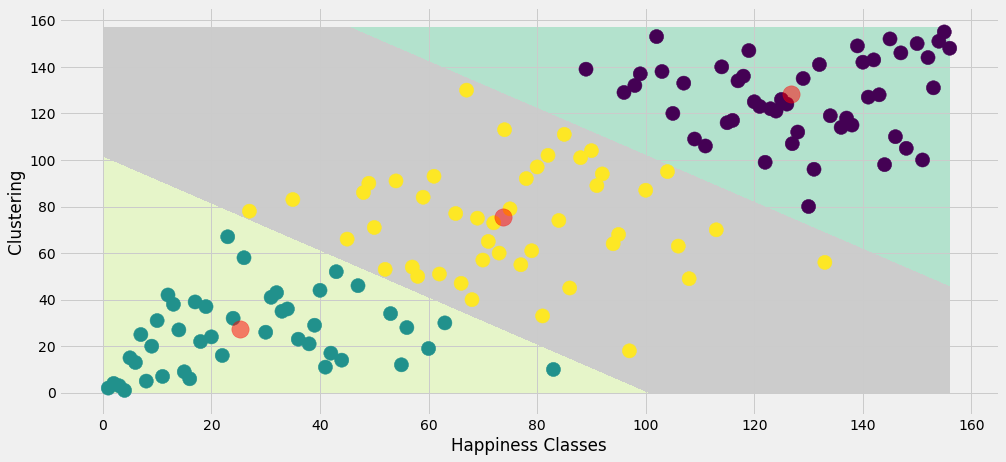}
         %\subcaption{Distribution of Happiness Score after prediction.}
         \label{rq8}
     \end{minipage}
     \hfill
     \caption{Visualization of Clustering Results based happiness scores}
   \end{figure}

%\begin{landscape}
\begin{table}[]
\centering
\caption{Empirical Results obtained from the Machine Learning Algorithms and
Clustering}
\label{tab:my-table}
\begin{tabular}{|l|llll|llll|}
\hline
 & \multicolumn{4}{l|}{\begin{tabular}[c]{@{}l@{}}Without \\ Clustering\end{tabular}} & \multicolumn{4}{l|}{Clustering} \\ \hline
 & LR & SVM & KNN & RF & LR & SVM & KNN & RF \\ \hline
F1-Score & 0.88 & 0.90 & 0.86 & 0.84 & 0.90 & 0.85 & 0.84 & 0.98 \\
Accuracy & 0.91 & 0.92 & 0.90 & 0.94 & 0.92 & 0.925 & 0.95& 0.96 \\
AUC & 0.96 & 0.94 & 0.96 & 0.96 & 0.96 & 0.97 & 0.97 & 0.98 \\
Train Score & 0.84 & 0.83 & 0.81 & 0.99 & 0.84 & 0.82 & 0.84 & 0.81 \\
Test Score & 0.875 & 0.77 & 0.84 & 0.875 & 0.87 & 0.70 & 0.71 & 0.82 \\ \hline
\end{tabular}
\end{table}
%\end{landscape}

\section{Experiments and Results}
This section highlights the empirical analysis and the results obtained.
\begin{itemize}

\item The source code for the experiment is available at "https://github.com/smlab-niser/2020happiness".The exploratory analysis of the survey data is listed in Table-2.

\item The model equation so obtained after applying regression analysis is -
 Happiness score = 0.0001289+ 1.000005Social +0.999869Health + 0.999912Freedom.
The average score of happiness is calculated to be 5.4. The predicted happiness score is depicted to be maximum for the Social and Health factor. Pearson's coefficient of correlation is found to be 0.65 for Social and 0.74 for Health. The Correlation coefficient of Australian students was 0.66. Table 3 and Table 5 depict the outcomes of empirical analysis upon applying regression. 
%R-square was found to be 0.986. It was found out that there were a few missing answers given by respondents denoting their uncertainty and inability to comprehend. Lyard\cite{11} underlined this as the soundness of the questionnaire in a survey. 

\item Table 4 captures the satisfaction in the university of the respondents in context with the statement-" Overall, I'm happy with my university life". %The statistics of Indian respondents were as follows- Very Happy(UR=15\%; TM, GSR, COE= 13\% each, CE+RF+WB+ECA=46\%).  Happy(TM=13.5\%; CE=13\%; WB=13\%; RF=16\%; UR+GSR+COE+ECA=44.5\% ). Neutral(RF+WB=13.5\%, ECA=12\% , COE=16\%,  TM+UR+CE+GSR+WB=45\%).  Sad(CE,GSR,TM=13\%; ECA=14.5\%; WB+COE+UR+RF=46.5\%). The happiness percentage of all the respondents was captured on a 4-point Likert scale(4-Very happy, 3-Happy, 2-Neutral, and 1-Sad)  based on the study of Chan and Mangeloja and Hirvonen. 86.6\% of the Finnish students were found to be more content compared to their Australian counterparts. 

\item Using STATA, the statistical outcomes so obtained were compared with those from the study of Chan\cite{18} and Esa\cite{17}. The predicted value of the score from the explanatory variable is computed and listed in Table-6. The explanatory variables namely, ECA and WB seem to be significant(p-value= 0.854 and 0.759); while CE and UR seem to be quite insignificant for our survey. GSR is also one of the predominant criteria for students' satisfaction, similar to the Finland study. the goodness of model fit was found at 51\% at 10\% CI.

\item Fig 3 depicts the exploratory analysis and the relationship amongst the predicted variables(Freedom, Health, Social), and the use of algorithms on the dataset based on the happiness classes.

\item Table 7 shows LR got the highest accuracy and F-score without clustering while RF+Clustering proved best for our model with the highest accuracy(0.89) and F-score(0.98). Fig. 5 depicts the visualization of Clustering algorithms on the dataset

\item Fig. 4a, 4c, 4e, depict the happiness score based on the predicted variables upon regression analysis. Fig. 4b, 4d, 4f, picturize the visualization of the model equation of predicted variables.

\item  Model scores are LR=71.56\%; SVM=71.41\%; Clustering=71.46\%; KNN=72.31\%; RF=74.79\%. The predicted happiness score turns out to be 5.2. RF+Clustering model got the least error=17.91\%(8 features). \% of variance explained=80.28. Mean Square Residual=0.24.

\item Philosophy was the happiest department and Sociology the saddest amongst all other  departments. Their average happiness scores resulted in 2.85 and 2.44 respectively. Sanskrit had a negligible impact on students' happiness with a minor variation. School of Women Studies, Odia, and Law had a wide fluctuation in happiness scores. 

\end{itemize}

\section{Conclusions and Future Scope}

The ensuing model from this study establishes a new state-of-the-art to prioritize only a cluster of university students, accumulate their opinions, introspect the happiness rank department-wise using machine learning and clustering paradigms. %Hence, we can unfold the following conclusions and reveal interesting points. A typical machine learning model incorporates k-fold cross validation of splitting the dataset into 30:70 or 20:80 ratio. And to the best of our knowledge the techniques used in our study is a bit unique.
The answer to the question of whether a Utkal university student was happy or not with his/her university life, is assertive in 73\% of the cases. This indicates that 69\% of students were satisfied with their university life. The research study of Esa\cite{17} had dis-aggregated TM into 3 categories(meeting deadlines, WB, and recreational time) and also the satisfaction of school work into 5( happy with marks, enjoying studies, interesting work, coping up, resources, and environment). Considering a 10\% CI, age differences and gender differences had an undersized impact. RF+Clustering got the highest accuracy(89\%) and F-score(0.98) and the least error(17.91\%), hence turning out to be the best for our study.
The future scope of our study can be: subsequently, we can proliferate the dataset by supplementing our online survey in other locally situated universities. A 7-scaled Likert scale could be used to make precise calculations. Experiments can be validated statistically using ANOVA and MANOVA test.

\end{document}